\documentclass[12pt]{article}

\author{Yu.~M.~Zinoviev
       \thanks{E-mail address: Yurii.Zinoviev@ihep.ru} \\
        {\it Institute for High Energy Physics} \\
        {\it Protvino, Moscow Region, 142280, Russia}}
\title{On electromagnetic interactions \\
for massive mixed symmetry field}

\date{}

\def\eptwo{\left\{ \phantom{|}^{\mu\nu}_{ab} \right\}}
\def\epthree{\left\{ \phantom{|}^{\mu\nu\alpha}_{abc} \right\}}
\def\epfour{\left\{ \phantom{|}^{\mu\nu\alpha\beta}_{abcd} \right\}}

\begin{document}

\maketitle

\begin{abstract}
In this paper we investigate electromagnetic interactions for simplest
massive mixed symmetry field. Using frame-like gauge invariant
formulation we extend Fradkin-Vasiliev procedure, initially proposed
for investigation of gravitational interactions for massless particles
in $AdS$ space, to the case of electromagnetic interactions for
massive particles leaving in $(A)dS$ space with arbitrary value of
cosmological constant including flat Minkowski space. At first, as an
illustration of general procedure, we re-derive our previous results
on
massive spin 2 electromagnetic interactions and then we apply this
procedure to massive mixed symmetry field. These two cases are just
the simplest representatives of two general class of fields, namely
completely symmetric and mixed symmetry ones, and it is clear that the
results obtained admit straightforward generalization to higher spins
as well.
\end{abstract}

\thispagestyle{empty}
\newpage
\setcounter{page}{1}

\section*{Introduction}

It has been known since a long time that it is not possible to
construct standard gravitational interaction for massless higher spin
$s \ge 5/2$ particles in flat Minkowski space (see \cite{BBS10} and
references therein). At the same time, it has been shown
\cite{FV87,FV87a} that this task indeed has a solution in $(A)dS$
space with non-zero cosmological term. The reason is that gauge
invariance, that turns out to be broken when one replaces ordinary
partial derivatives by the gravitational covariant ones, could be
restored with the introduction of higher derivative corrections
containing gauge invariant Riemann tensor. These corrections have
coefficients proportional to inverse powers of cosmological constant
so that such theories do not have naive flat limit. However it is
perfectly possible, for cubic vertices, to have a limit where both
cosmological term and gravitational coupling constant simultaneously
go to zero in such a way that only interactions with highest number of
derivatives survive \cite{Zin08,BLS08}.

Besides gravitational interaction one more classical and important
test for any higher spin theory is electromagnetic interaction. The
problem of switching on such interaction for massless higher spin
particles looks very similar to the problem with gravitational
interactions. Namely, if one replaces ordinary partial derivatives by
the gauge covariant ones the resulting Lagrangian loses its gauge
invariance and this non-invariance (arising due to non-commutativity
of covariant derivatives) is proportional  to field strength of vector
field. In this, for the massless fields with $s \ge 3/2$ in flat
Minkowski space  there is no possibility to restore gauge invariance
by adding non-minimal terms to Lagrangian and/or modifying gauge
transformations. But such restoration becomes possible if one goes to
$(A)dS$ space with non-zero cosmological constant. By the same reason,
as in the gravitational case, such theories do not have naive flat
limit, but it is possible to consider a limit where both cosmological
constant and electric charge simultaneously go to zero so that only
highest derivative non-minimal terms survive \cite{BLS08,Zin08a}. 

It is natural to suggest that in any realistic higher spin theory
(like in superstring) most of higher spin particles must be massive
and their gauge symmetries spontaneously broken. As is well known, for
massive higher spin particles any attempt to switch on standard
minimal gravitational or electromagnetic interactions spoils a
consistency of the theory leading first of all to appearance of 
non-physical degrees of freedom and/or non-causality. But having in
our disposal mass $m$ as a dimensionfull parameter even in a flat
Minkowski space we can try to restore consistency of the theory by
adding to Lagrangian non-minimal terms containing the linearized
Riemann tensor (e/m field strength). Naturally such terms will have
coefficients proportional to inverse powers of mass $m$ so that the
theory will not have naive massless limit. However, it is natural to
suggest that there exists a limit where both mass and gravitational
coupling constant (electric charge) simultaneously go to zero so that
only some interactions containing Riemann tensor (e/m field strength)
survive. Note that such a picture agrees with the results obtained
from (super)strings \cite{Tar10,ST10,Sch10}.

As is well known, in four dimensions to describe all possible higher
spins it is enough to consider completely symmetric (spin-)tensors
only. But in dimensions greater than four it is necessary to consider
mixed symmetry (spin-)tensors as well. There exists a number of works
devoted to investigation of possible interactions for such fields
\cite{BCNS02,BBH02,BCCSS03,BCCSS03a,Sal03,BC04,BBC04,CS07,BCNS07,BCCS07,
BCS09,BCDISS09}, but till now most of them produce so called "no-go"
results (note however light-cone results \cite{FM95} and recent work
\cite{Alk10}). 

In this paper we investigate electromagnetic interactions for simplest
massive mixed symmetry field ("hook"). In this, we extend
Fradkin-Vasiliev procedure, initially proposed for investigation of
gravitational interactions for massless particles in $AdS$ space and
then successfully applied to other massless higher spin actions 
\cite{Vas96,Vas01,AV02,Alk10}, to the case of electromagnetic
interactions for massive particles leaving in $(A)dS$ space with
arbitrary value of cosmological constant including flat Minkowski
space. One of the essential ingredients of the procedure is the
frame-like formalism, initially introduced for completely symmetric
(spin-)tensors \cite{Vas80,LV88,Vas88} (see also \cite{Zin08b,PV10})
and then extended to the case of massless 
\cite{Zin03,ASV03,Alk03,ASV05,ASV06,Skv08a,Skv08,SZ10} and massive
\cite{Zin03a,Zin08c,Zin09b,Zin09c} mixed symmetry (spin-)tensors.
Briefly, the procedure can be described as follows.
\begin{itemize}
\item Take as input frame-like gauge invariant formulation for massive
particle.
\item Using explicit form of gauge transformations construct for all
fields (both physical and auxiliary) gauge invariant objects which in
what follows we will call "curvatures" though there will be one, two
and three forms among them.
\item Rewrite free Lagrangian as an expression quadratic in these
curvatures. Note that in general even if one requires that all higher
derivatives terms were absent and kinetic terms were diagonal one
still faces some ambiguity in the choice of coefficients. 
\item Find deformations of these curvatures supplemented with
appropriate corrections to gauge transformations such that their gauge
variations be proportional to free curvatures. In the case of
electromagnetic interactions the only non-trivial task is to  find
deformation for electromagnetic field strength while all we have to
do with other curvatures is to replace $AdS$ covariant derivatives
with the fully covariant ones.
\item In the free Lagrangian replace free curvatures with the deformed
ones and adjust free parameters so that all variations vanish
on-shell.
Note that at least in some cases this resolves ambiguity with the
parameters in the free Lagrangian as well.
\item Determine corrections to gauge transformations such that all
gauge variations vanish off-shell.
\end{itemize}

The plan of the paper is simple. In Section 1, as an illustration of
general procedure, we re-derive our previous results on massive spin 2
electromagnetic interactions \cite{Zin09}. All physical results are
the same, but derivation becomes much more simple and transparent.
Then in Section 2 we apply this procedure to the electromagnetic
interactions for simplest massive mixed symmetry tensor.

{\bf Notations and conventions.} We work in $(A)dS$ space
with $d \ge 4$ dimensions. We will use notation $e_\mu{}^a$ for
background (non-dynamical) frame of $(A)dS$ space and $D_\mu$ for
$(A)dS$ covariant derivatives normalized so that
$$
[ D_\mu, D_\nu ] \xi^a = - \kappa e_{[\mu}{}^a \xi_{\nu]}, \qquad
\kappa = \frac{2\lambda}{(d-1)(d-2)}
$$
We use Greek letters for world indices and Latin letters for local
ones. Surely, using frame $e_\mu{}^a$ and its inverse $e^\mu{}_a$ one
can freely convert world indices into local ones and vice-verse and we
indeed will use such conversion whenever convenient. But separation of
world and local indices plays very important role in a frame-like
formalism. In particular, all terms in the Lagrangians can be written
as a product of forms, i.e. as expressions completely antisymmetric on
world indices and this property greatly simplifies all calculations.
For that purpose we will often use notations $\eptwo = e^\mu{}_a
e^\nu_b - e^\nu{}_a e^\mu{}_b$ and so on.

\section{Massive spin 2}

In this section, as an illustration of general procedure, we re-derive
our previous results on spin 2 electromagnetic interactions
\cite{Zin08a,Zin09}.

\subsection{Lagrangian and gauge transformations}

Frame-like gauge invariant formulations for massive spin 2 particles
\cite{Zin03a,Zin08b} requires three pairs of physical and auxiliary
fields: ($\omega_\mu{}^{ab}$, $f_\mu{}^a$), ($B^{ab}$, $B_\mu$) and
($\pi^a$, $\varphi$). The free Lagrangian in the constant curvature
space with arbitrary value of cosmological constant has the form:
\begin{eqnarray}
{\cal L}_0 &=& \frac{1}{2} \eptwo \omega_\mu{}^{ac} \omega_\nu{}^{bc}
- \frac{1}{2} \epthree \omega_\mu{}^{ab} D_\nu f_\alpha{}^c + 
\frac{1}{4} B_{ab}{}^2 - \frac{1}{2} \eptwo  B^{ab} D_\mu B_\nu -
\nonumber \\
 && - \frac{1}{2} \pi_a{}^2 + e^\mu{}_a \pi^a D_\mu \varphi + m [
\eptwo \omega_\mu{}^{ab} B_\nu + e^\mu{}_a B^{ab} f_\mu{}^b
] - \tilde{M} e^\mu{}_a \pi^a B_\mu + \nonumber \\
 && + M^2 \eptwo f_\mu{}^a f_\nu{}^b - m\tilde{M} e^\mu{}_a f_\mu{}^a
\varphi + \frac{d}{(d-2)} m^2 \varphi^2
\end{eqnarray}
Here $M^2 = m^2 - \frac{\kappa(d-2)}{2}$, $\tilde{M} = 
2\sqrt{\frac{(d-1)}{(d-2)}}M$.

This Lagrangian is invariant under the following gauge
transformations:
\begin{eqnarray}
\delta_0 f_\mu{}^a &=& D_\mu \xi^a + \eta_\mu{}^a + \frac{2m}{(d-2)}
e_\mu{}^a \Lambda, \qquad \delta_0 \omega_\mu{}^{ab} = D_\mu \eta^{ab}
- \frac{2M^2}{(d-2)} e_\mu{}^{[a} \xi^{b]} \nonumber \\
\delta_0 B_\mu &=& D_\mu \Lambda + m \xi_\mu, \qquad
\delta_0 B^{ab} = - 2m \eta^{ab}, \qquad
\delta_0 \varphi = \tilde{M} \Lambda, \qquad
\delta_0 \pi^a = - m\tilde{M} \xi^a
\end{eqnarray}

Note that in a de Sitter space ($\kappa > 0$) there exists so called
partially massless limit \cite{DW01,DW01a,DW01c,Zin01,SV06,DW06} where
spin 0 component completely decouples leaving us with the Lagrangian
\begin{eqnarray}
{\cal L}_0 &=& \frac{1}{2} \eptwo \omega_\mu{}^{ac} \omega_\nu{}^{bc}
- \frac{1}{2} \epthree \omega_\mu{}^{ab} D_\nu f_\alpha{}^c + 
\frac{1}{4} B_{ab}{}^2 - \frac{1}{2} \eptwo  B^{ab} D_\mu B_\nu +
\nonumber \\
 && + m [ \eptwo \omega_\mu{}^{ab} B_\nu + e^\mu{}_a B^{ab} f_\mu{}^b
] 
\end{eqnarray}
which is invariant under the following gauge transformations
\begin{eqnarray}
\delta_0 f_\mu{}^a &=& D_\mu \xi^a + \eta_\mu{}^a + \frac{2m}{(d-2)}
e_\mu{}^a \Lambda, \qquad \delta_0 \omega_\mu{}^{ab} = D_\mu \eta^{ab}
\nonumber \\
\delta_0 B_\mu &=& D_\mu \Lambda + m \xi_\mu, \qquad
\delta_0 B^{ab} = - 2m \eta^{ab}
\end{eqnarray}

\subsection{Gauge invariant objects}

For all six fields (both physical and auxiliary) we can construct
corresponding gauge invariant object ("curvature"):
\begin{eqnarray}
{\cal F}_{\mu\nu}{}^{ab} &=& D_{[\mu} \omega_{\nu]}{}^{ab} -
\frac{m}{(d-2)} e_{[\mu}{}^{[a} B_{\nu]}{}^{b]} - \frac{2M^2}{(d-2)} 
e_{[\mu}{}^{[a}  f_{\nu]}{}^{b]} + \frac{2m\tilde{M}}{(d-1)(d-2)}
e_{[\mu}{}^a e_{\nu]}{}^b \varphi \nonumber \\
T_{\mu\nu}{}^a &=& D_{[\mu} f_{\nu]}{}^a - \omega_{[\mu,\nu]}{}^a +
\frac{2m}{(d-2)} e_{[\mu}{}^a B_{\nu]} \nonumber \\
{\cal B}_\mu{}^{ab} &=& D_\mu B^{ab} + 2m \omega_\mu{}^{ab} -
\frac{\tilde{M}}{(d-1)} e_\mu{}^{[a} \pi^{b]} \nonumber \\
{\cal B}_{\mu\nu} &=& D_{[\mu} B_{\nu]} - B_{\mu\nu} - m
f_{[\mu,\nu]} \\
\Pi_\mu{}^a &=& D_\mu \pi^a + \frac{\tilde{M}}{2} B_\mu{}^a
+ m\tilde{M} f_\mu{}^a - \frac{2m^2}{(d-2)} e_\mu{}^a \varphi
\nonumber \\
\Phi_\mu &=& D_\mu \varphi - \pi_\mu - \tilde{M} B_\mu \nonumber
\end{eqnarray}

As it has been shown in \cite{PV10} for arbitrary spin massive
particles the free Lagrangian can be rewritten as expression quadratic
in these gauge invariant curvatures. For the case at hands the most
general such Lagrangian can be written as follows:
\begin{eqnarray}
{\cal L}_0 &=& a_1 \epfour {\cal F}_{\mu\nu}{}^{ab}
{\cal F}_{\alpha\beta}{}^{cd} + a_2 \epthree {\cal F}_{\mu\nu}{}^{ab}
\Pi_\alpha{}^c + a_3 \eptwo {\cal B}_\mu{}^{ac} {\cal B}_\nu{}^{bc} +
a_4 \eptwo \Pi_\mu{}^a \Pi_\nu{}^b + \nonumber \\
 && + b_1 \epthree T_{\mu\nu}{}^a {\cal B}_\alpha{}^{bc} + b_2 \eptwo
{\cal B}_\mu{}^{ab} \Phi_\nu
\end{eqnarray}
Note that even if we require that higher derivatives terms were absent
and kinetic terms were diagonal there still exists some ambiguity in
the choice of coefficients. In what follows (see below) we will use
the following simple concrete choice:
\begin{equation}
{\cal L}_0 = a_1 \epfour {\cal F}_{\mu\nu}{}^{ab}
{\cal F}_{\alpha\beta}{}^{cd} + a_3 \eptwo {\cal B}_\mu{}^{ac} 
{\cal B}_\nu{}^{bc} + b_2 \eptwo {\cal B}_\mu{}^{ab} \Phi_\nu
\end{equation}
where
$$
\tilde{a}_1 = \frac{16(d-3)}{(d-2)} a_1 = \frac{1}{4M^2}, \qquad
a_3 = \frac{1}{8M^2}, \qquad b_2 = - \frac{1}{2\tilde{M}}
$$
Note that up to normalization it is just the Ponomarev-Vasiliev
Lagrangian (4.20) in \cite{PV10}.

\subsection{Deformations}

Now let us turn to the electromagnetic interactions. We prefer to work
with real fields, so from now on all components of massive spin 2 will
become doublets $f_\mu{}^a \to f_\mu{}^{a,i}$, $i=1,2$ and so on. Our
first task --- to find appropriate deformations of all gauge invariant
curvatures given above as well of electromagnetic field strength. As
for massive spin 2 curvatures, the answer appears to be very simple:
all we have to do is to replace $(A)dS$ covariant derivatives by fully
covariant ones:
$$
D_\mu \quad \Longrightarrow \quad \nabla_\mu{}^{ij} = \delta^{ij}
D_\mu + e_0 \varepsilon^{ij} A_\mu
$$
where $e_0$ --- electric charge. As for the electromagnetic field
strength, having in our disposal explicit expressions for curvatures
and gauge transformations it is not hard to find desired result.

Let us consider the following ansatz for such deformation:
$$
\hat{F}_{\mu\nu} = F_{\mu\nu} + a_0 \varepsilon^{ij} [
\omega_{[\mu}{}^{ab,i} \omega_{\nu]}{}^{ab,j} + \alpha_1
B_{[\mu}{}^{a,i} B_{\nu]}{}^{a,j} + \alpha_2 f_{[\mu}{}^{a,i} 
f_{\nu]}{}^{a,j} + \alpha_3 B_{[\mu}{}^i B_{\nu]}{}^j + \alpha_4
B_{\mu\nu}{}^i \varphi^j ]
$$
Its variation under $\eta^{ab}$ gauge transformations has the form:
$$
\delta \hat{F}_{\mu\nu} = 2a_0 \varepsilon^{ij} [
- D_{[\mu} \eta^{ab,j} \omega_{\nu]}{}^{ab,i} 
- 2 m \alpha_1 B_{[\mu}{}^{a,i} \eta_{\nu]}{}^{a,j} 
+ \alpha_2 f_{[\mu}{}^{a,i} \eta_{\nu]}{}^{a,j}
+ m \alpha_4 \varphi^i \eta_{\mu\nu}{}^j ]
$$
Now we introduce correction to e/m field gauge transformations:
$$
\delta A_\mu = 2a_0 \varepsilon^{ij} \omega_\mu{}^{ab,i} \eta^{ab,j}
$$
As a result we obtain:
$$
\delta \hat{F}_{\mu\nu} = 2a_0 \varepsilon^{ij} [
D_{[\mu} \omega_{\nu]}{}^{ab,i} \eta^{ab,j}
- 2 m \alpha_1 B_{[\mu}{}^{a,i} \eta_{\nu]}{}^{a,j} 
+ \alpha_2 f_{[\mu}{}^{a,i} \eta_{\nu]}{}^{a,j}
+ m \alpha_4 \varphi^i \eta_{\mu\nu}{}^j ]
$$
Comparing this expression with
\begin{eqnarray*}
2a_0 \varepsilon^{ij} {\cal F}_{\mu\nu}{}^{ab,i} \eta^{ab,j} &=&
2a_0 \varepsilon^{ij} [ D_{[\mu} \omega_{\nu]}{}^{ab,i} -
\frac{2m}{(d-2)} e_{[\mu}{}^a B_{\nu]}{}^{b,i} -
\frac{4M^2}{(d-2)} e_{[\mu}{}^a f_{\nu]}{}^{b,i} + \\
 && \qquad \quad + \frac{2m\tilde{M}}{(d-1)(d-2)} e_{[\mu}{}^a
e_{\nu]}{}^b \varphi^i ] \eta^{ab,j}
\end{eqnarray*}
we see that we have to put
$$
\alpha_1 = - \frac{1}{(d-2)}, \qquad
\alpha_2 = \frac{4M^2}{(d-2)}, \qquad
\alpha_4 = \frac{4\tilde{M}}{(d-1)(d-2)}
$$
Similarly, variation under $\xi^a$ transformations:
$$
\delta \hat{F}_{\mu\nu} = 2a_0 \varepsilon^{ij} [ - \alpha_2 D_{[\mu}
\xi^{a,j} f_{\nu]}{}^{a,i} - \frac{4M^2}{(d-2)} 
\omega_{[\mu,\nu]}{}^{a,i} \xi^{a,j} + m\alpha_3 B_{[\mu}{}^i
\xi_{\nu]}{}^j ]
$$
together with corresponding correction
$$
\delta A_\mu = 2a_0\alpha_2 \varepsilon^{ij} f_\mu{}^{a,i} \xi^{a,j}
$$
produce
$$
\delta \hat{F}_{\mu\nu} = 2a_0 \varepsilon^{ij} [ \alpha_2 \xi^{a,j}
D_{[\mu} f_{\nu]}{}^{a,i} - \frac{4M^2}{(d-2)} 
\omega_{[\mu,\nu]}{}^{a,i} \xi^{a,j} + m\alpha_3 B_{[\mu}{}^i
\xi_{\nu]}{}^j ]
$$
Comparing this expression with
$$
2a_0 \alpha_2 \varepsilon^{ij} T_{\mu\nu}{}^{a,i} \xi^{a,j} =
2a_0 \alpha_2 \varepsilon^{ij} [ D_{[\mu} f_{\nu]}{}^{a,i} -
\omega_{[\mu,\nu]}{}^{a,i} + \frac{2m}{(d-2)} e_{[\mu}{}^a 
B_{\nu]}{}^i ] \xi^{a,j}
$$
we get
$$
\alpha_3 = - \frac{8M^2}{(d-2)^2}
$$
All the unknown coefficients are already fixed, but as a useful check
one can easily see that variation under $\Lambda$ transformations
$$
\delta \hat{F}_{\mu\nu} = a_0 \varepsilon^{ij} [ - 2\alpha_3 D_{[\mu}
\Lambda^j B_{\nu]}{}^i + \frac{4m\alpha_2}{(d-2)} f_{[\mu,\nu]}{}^i
\Lambda^j + \tilde{M} \alpha_4 B_{\mu\nu}{}^i \Lambda^j ]
$$
together with
$$
\delta A_\mu = 2\alpha_3 a_0 \varepsilon^{ij} B_\mu{}^i \Lambda^j
$$
give the desired result:
$$
\delta \hat{F}_{\mu\nu} = a_0 \varepsilon^{ij} [ 2\alpha_3 D_{[\mu}
 B_{\nu]}{}^i \Lambda^j + \frac{4m\alpha_2}{(d-2)} f_{[\mu,\nu]}{}^i
\Lambda^j + \tilde{M} \alpha_4 B_{\mu\nu}{}^i \Lambda^j ] =
- \frac{16M^2a_0}{(d-2)^2} \varepsilon^{ij} {\cal B}_{\mu\nu}{}^i
\Lambda^j
$$
Collecting all pieces together, we obtain finally
\begin{eqnarray}
\hat{F}_{\mu\nu} &=& F_{\mu\nu} + a_0 \varepsilon^{ij} [ 
\omega_{[\mu}{}^{ab,i} \omega_{\nu]}{}^{ab,j}
- \frac{1}{(d-2)} B_{[\mu}{}^{a,i} B_{\nu]}{}^{a,j} 
+ \frac{4M^2}{(d-2)} f_{[\mu}{}^{a,i} f_{\nu]}{}^{a,j} - \nonumber \\
 && \qquad \qquad - \frac{8M^2}{(d-2)^2} B_{[\mu}{}^i B_{\nu]}{}^j
+ \frac{4\tilde{M}}{(d-1)(d-2)} B_{\mu\nu}{}^i \varphi^j ]
\end{eqnarray}
In this, under massive spin 2 gauge transformations
supplemented with the following corrections to e/m field gauge
transformations:
\begin{equation}
\delta A_\mu = 2a_0 \varepsilon^{ij} [ \omega_\mu{}^{ab,i}
\eta^{ab,j} + \frac{4M^2}{(d-2)} f_\mu{}^{a,i} \xi^{a,j} -
\frac{8M^2}{(d-2)^2} B_\mu{}^i \Lambda^j ]
\end{equation}
this tensor transforms as:
\begin{equation}
\delta \hat{F}_{\mu\nu} = 2 a_0 \varepsilon^{ij} [ 
{\cal F}_{\mu\nu}{}^{ab,i} \eta^{ab,j}
+ \frac{4M^2}{(d-2)} T_{\mu\nu}{}^{a,i} \xi^{a,j} -
\frac{8M^2}{(d-2)^2} {\cal B}_{\mu\nu}{}^i \Lambda^j ]
\end{equation}

Now, following general procedure, we consider Lagrangian:
\begin{equation}
{\cal L}_0 = a_1 \epfour \hat{\cal F}_{\mu\nu}{}^{ab}
\hat{\cal F}_{\alpha\beta}{}^{cd} + a_3 \eptwo \hat{\cal B}_\mu{}^{ac}
\hat{\cal B}_\nu{}^{bc} + b_2 \eptwo \hat{\cal B}_\mu{}^{ab}
\hat\Phi_\nu - \frac{1}{4} \hat{F}_{\mu\nu}{}^2
\end{equation}

\subsection{Gauge invariance}

Now we have to calculate variations of this Lagrangian in the first
non-trivial approximation and to adjust parameters so that all such
variations vanish on-shell. This in turn would imply that they can be
compensated by appropriate corrections to gauge transformations.

$\eta^{ab}$ {\bf transformations}. In this case we obtain the
following variations of our Lagrangian:
\begin{eqnarray}
\delta_\eta {\cal L} &=& - 4 a_1 e_0 \varepsilon^{ij} \eptwo [ 4 
{\cal F}_{\mu\nu}{}^{ac,i} \eta^{bd,j} - {\cal F}_{\mu\nu}{}^{ab,i}
\eta^{cd,j} ] F^{cd} + \nonumber \\
 && + \varepsilon^{ij} {\cal F}_{\mu\nu}{}^{ab,i} [ 8a_1e_0
\eta^{\mu\nu,j} F^{ab} - a_0 F^{\mu\nu} \eta^{ab,j} ] 
\end{eqnarray}
Terms in the first line clearly vanish on-shell being proportional to
free $f_\mu{}^a$ equations. Now, using explicit expressions for free
curvatures it is not hard to check that the following identity holds:
\begin{equation}
{\cal F}_{[\mu\nu,\alpha]}{}^a = - D_{[\mu} T_{\nu\alpha]}{}^a -
\frac{2m}{(d-2)} e_{[\mu}{}^a {\cal B}_{\nu\alpha]}
\end{equation}
Thus ${\cal F}_{[\mu\nu,\alpha]}{}^a$ vanish on-shell and as a
consequence we obtain
$$
{\cal F}_{[\mu\nu,\alpha]}{}^a = 0 \qquad \Longrightarrow \qquad
{\cal F}_{ab,cd} = {\cal F}_{cd,ab}
$$
So if we set $a_0 = 8a_1e_0$ then all $\eta^{ab}$ variations vanish on
shell and can be compensated by appropriate corrections to gauge
transformations. For simplicity we restrict ourselves with the
corrections to gauge transformations for physical fields only which in
this case have the form:
\begin{equation}
\delta_1 f_\mu{}^{a,i} = 4 a_0 \varepsilon^{ij} [ F_\mu{}^b
\eta^{ab,j} - \frac{1}{2(d-2)} e_\mu{}^a (F \eta)^j ]
\end{equation}

$\xi^a$ {\bf transformations}. Here we obtain simply
$$
\delta_\xi {\cal L} =  - \frac{4M^2}{(d-2)} a_0
\varepsilon^{ij} F^{\mu\nu} T_{\mu\nu}{}^{a,i} \xi^{a,j}
$$
This term vanish on-shell, in this it can be compensated
by corresponding corrections to $\omega_\mu{}^{ab}$ transformations.

$\Lambda$ {\bf transformations}. Here we get
$$
\delta_\Lambda {\cal L} = - \frac{8M^2}{(d-2)^2} a_0
\varepsilon^{ij} F^{\mu\nu} {\cal B}_{\mu\nu}{}^i \Lambda^j
$$
This term also vanish on-shell and can be compensated
by corresponding corrections to $B^{ab}$ transformations.

\subsection{Results}

Thus we have seen that the following Lagrangian
\begin{equation}
{\cal L}_0 = a_1 \epfour \hat{\cal F}_{\mu\nu}{}^{ab}
\hat{\cal F}_{\alpha\beta}{}^{cd} + a_3 \eptwo \hat{\cal B}_\mu{}^{ac}
\hat{\cal B}_\nu{}^{bc} + b_2 \eptwo \hat{\cal B}_\mu{}^{ab}
\hat\Phi_\nu - \frac{1}{4} \hat{F}_{\mu\nu}{}^2
\end{equation}
is indeed gauge invariant in the linear approximation provided we
introduce corresponding corrections to gauge transformations
\begin{eqnarray}
\delta A_\mu &=& 2a_0 \varepsilon^{ij}[ \omega_\mu{}^{ab,i}
\eta^{ab,j} + \frac{4M^2}{(d-2)} f_\mu{}^{a,i} \xi^{a,j} - 
\frac{8M^2}{(d-2)^2} B_\mu{}^i \Lambda^j ] \nonumber \\
\delta f_\mu{}^{a,i} &=& 4 a_0 \varepsilon^{ij} [ F_\mu{}^b
\eta^{ab,j} - \frac{1}{2(d-2)} e_\mu{}^a (F \eta)^j ]
\end{eqnarray}

Note that in agreement with our previous results in \cite{Zin09} it is
impossible to take partially massless limit $M \to 0$ in de Sitter
space without switching off minimal e/m interactions. At the same
time, nothing prevents us from considering massless limit in anti de
Sitter space $m=0$, $M^2 = - \frac{\kappa(d-2)}{2}$. In this limit we
obtain simple Lagrangian
\begin{equation}
{\cal L} = a_1 \epfour {\cal F}_{\mu\nu}{}^{ab} 
{\cal F}_{\alpha\beta}{}^{cd} - \frac{1}{4} \hat{F}_{\mu\nu}{}^2
\end{equation}
where
\begin{eqnarray}
{\cal F}_{\mu\nu}{}^{ab} &=&  \nabla_{[\mu} \omega_{\nu]}{}^{ab} -
\frac{2M^2}{(d-2)} e_{[\mu}{}^{[a} f_{\nu]}{}^{b]} ]  \nonumber \\
\hat{F}_{\mu\nu} &=& F_{\mu\nu} + a_0 \varepsilon^{ij} [ 
\omega_{[\mu}{}^{ab,i} \omega_{\nu]}{}^{ab,j} + \frac{4M^2}{(d-2)}
f_{[\mu}{}^{a,i} f_{\nu]}{}^{a,j} ]
\end{eqnarray}

\section{Mixed symmetry tensor}

In this section we apply the same procedure to construct
electromagnetic interactions for simplest massive mixed symmetry
tensor ("hook").

\subsection{Lagrangian and gauge transformations}

In this case frame-like gauge invariant description
\cite{Zin03a,Zin08c} requires four pairs of physical and auxiliary
fields: $(\Omega_\mu{}^{abc}, \Phi_{\mu\nu}{}^a)$,
$(\omega_\mu{}^{ab}, h_\mu{}^a)$, $(C^{abc}, C_{\mu\nu})$ and
$(B^{ab}, B_\mu)$. Free Lagrangian has the form:
\begin{eqnarray}
{\cal L}_0 &=& - \frac{3}{4} \left\{ \phantom{|}^{\mu\nu}_{ab}
\right\} \Omega_\mu{}^{acd} \Omega_\nu{}^{bcd} + \frac{1}{4} \left\{
\phantom{|}^{\mu\nu\alpha\beta}_{abcd} \right\} \Omega_\mu{}^{abc}
D_\nu \Phi_{\alpha\beta}{}^d + \nonumber \\
 && + \frac{1}{2} \left\{ \phantom{|}^{\mu\nu}_{ab} \right\}
\omega_\mu{}^{ac} \omega_\nu{}^{bc} - \frac{1}{2} \left\{
\phantom{|}^{\mu\nu\alpha}_{abc} \right\} \omega_\mu{}^{ab} D_\nu
f_\alpha{}^c \nonumber - \\
 && - \frac{1}{6} C_{abc}{}^2 + \frac{1}{6} \left\{
\phantom{|}^{\mu\nu\alpha}_{abc} \right\} C^{abc} D_\mu
C_{\nu\alpha} + \frac{1}{4} B_{ab}{}^2 - \frac{1}{2} \left\{
\phantom{|}^{\mu\nu}_{ab} \right\} B^{ab} D_\mu B_\nu + \nonumber \\
 && + m_1 [ \left\{ \phantom{|}^{\mu\nu}_{ab} \right\}
\Omega_\mu{}^{abc} f_\nu{}^c + \left\{
\phantom{|}^{\mu\nu\alpha}_{abc} \right\}
\omega_\mu{}^{ab} \Phi_{\nu\alpha}{}^c ] + \nonumber \\
 && + m_2 [ \left\{ \phantom{|}^{\mu\nu\alpha}_{abc} \right\}
\Omega_\mu{}^{abc} C_{\nu\alpha} + \left\{
\phantom{|}^{\mu\nu}_{ab} \right\} C^{abc} 
\Phi_{\mu\nu}{}^c] + \nonumber \\
 && + 2 \tilde{m}_2 [ \left\{ \phantom{|}^{\mu\nu}_{ab} \right\}
\omega_\mu{}^{ab} B_\nu + \left\{\phantom{|}^\mu_a \right\} B^{ab}
f_\mu{}^b] + \tilde{m}_1 \left\{ \phantom{|}^{\mu\nu}_{ab} \right\}
B^{ab} C_{\mu\nu} 
\end{eqnarray}
parameters $m_{1,2}$ satisfy a relation
$$
8 m_1{}^2 - 24 m_2{}^2 = - 3 (d-3) \kappa
$$
while $\tilde{m}_{1,2} = \sqrt{\frac{(d-2)}{(d-3)}}m_{1,2}$. This
Lagrangian is invariant under the following set of gauge
transformations:
\begin{eqnarray}
\delta_0 \Phi_{\mu\nu}{}^a &=& D_{[\mu} z_{\nu]}{}^a + 
\eta_{\mu\nu}{}^a + \frac{2m_1}{3(d-3)} e_{[\mu}{}^a
\xi_{\nu]} + \frac{4m_2}{(d-3)} e_{[\mu}{}^a \zeta_{\nu]} \nonumber \\
\delta_0 \Omega_\mu{}^{abc} &=& D_\mu \eta^{abc} + \frac{4m_1}{3(d-3)}
e_\mu{}^{[a} \eta^{bc]} \nonumber \\
\delta_0 f_\mu{}^a &=& D_\mu \xi^a + \eta_\mu{}^a + 4 m_1 z_\mu{}^a +
\frac{4\tilde{m}_2}{(d-2)} e_\mu{}^a \Lambda \\ 
\delta_0 \omega_\mu{}^{ab} &=& D_\mu \eta^{ab} - 2 m_1 \eta_\mu{}^{ab}
\nonumber \\
\delta_0 C_{\mu\nu} &=& D_{[\mu} \zeta_{\nu]} - 2 m_2 z_{[\mu\nu]},
\qquad \delta_1 C^{abc} = 6 m_2 \eta^{abc} \nonumber \\
\delta_0 B_\mu &=& D_\mu \Lambda + 2 \tilde{m}_2 \xi_\mu + 4
\tilde{m}_1 \zeta_\mu, \qquad \delta_0 B^{ab} = - 4 \tilde{m}_2
\eta^{ab} \nonumber
\end{eqnarray}

As the relation on the parameters $m_{1,2}$ clearly shows for non-zero
values of cosmological constant $\kappa$ it is not possible to set
both $m_1$ and $m_2$ equal to zero simultaneously. In $AdS$ space
($\kappa < 0$) one can set $m_2 = 0$. In this, the whole system
decomposes into two disconnected subsystems. One of them with the
Lagrangian and gauge transformations:
\begin{eqnarray}
{\cal L}_0 &=& - \frac{3}{4} \left\{ \phantom{|}^{\mu\nu}_{ab}
\right\} \Omega_\mu{}^{acd} \Omega_\nu{}^{bcd} + \frac{1}{4} \left\{
\phantom{|}^{\mu\nu\alpha\beta}_{abcd} \right\} \Omega_\mu{}^{abc}
D_\nu \Phi_{\alpha\beta}{}^d + \nonumber \\
 && + \frac{1}{2} \left\{ \phantom{|}^{\mu\nu}_{ab} \right\}
\omega_\mu{}^{ac} \omega_\nu{}^{bc} - \frac{1}{2} \left\{
\phantom{|}^{\mu\nu\alpha}_{abc} \right\} \omega_\mu{}^{ab} D_\nu
f_\alpha{}^c \nonumber - \\
 && + m_1 [ \left\{ \phantom{|}^{\mu\nu}_{ab} \right\}
\Omega_\mu{}^{abc} f_\nu{}^c + \left\{
\phantom{|}^{\mu\nu\alpha}_{abc} \right\}
\omega_\mu{}^{ab} \Phi_{\nu\alpha}{}^c ] 
\end{eqnarray}
\begin{eqnarray}
\delta_0 \Phi_{\mu\nu}{}^a &=& D_{[\mu} z_{\nu]}{}^a + 
\eta_{\mu\nu}{}^a + \frac{2m_1}{3(d-3)} e_{[\mu}{}^a
\xi_{\nu]} \nonumber \\
\delta_0 \Omega_\mu{}^{abc} &=& D_\mu \eta^{abc} + \frac{4m_1}{3(d-3)}
e_\mu{}^{[a} \eta^{bc]} \nonumber \\
\delta_0 f_\mu{}^a &=& D_\mu \xi^a + \eta_\mu{}^a + 4 m_1 z_\mu{}^a \\
\delta_0 \omega_\mu{}^{ab} &=& D_\mu \eta^{ab} - 2 m_1 \eta_\mu{}^{ab}
\nonumber
\end{eqnarray}
corresponds to massless representation of $AdS$ group (which differs
from that of Poincare group \cite{BMV00}), while the other one just
gives gauge invariant description of massive antisymmetric second rank
tensor. In turn, in $dS$ space one can set $m_1 = 0$. In this case the
whole system also decomposes into two disconnected subsystems. One of
them with the Lagrangian and gauge transformations:
\begin{eqnarray}
{\cal L}_0 &=& - \frac{3}{4} \left\{ \phantom{|}^{\mu\nu}_{ab}
\right\} \Omega_\mu{}^{acd} \Omega_\nu{}^{bcd} + \frac{1}{4} \left\{
\phantom{|}^{\mu\nu\alpha\beta}_{abcd} \right\} \Omega_\mu{}^{abc}
D_\nu \Phi_{\alpha\beta}{}^d + \nonumber \\
 && - \frac{1}{6} C_{abc}{}^2 + \frac{1}{6} \left\{
\phantom{|}^{\mu\nu\alpha}_{abc} \right\} C^{abc} D_\mu
C_{\nu\alpha} + \nonumber \\
 && + m_2 [ \left\{ \phantom{|}^{\mu\nu\alpha}_{abc} \right\}
\Omega_\mu{}^{abc} C_{\nu\alpha} + \left\{
\phantom{|}^{\mu\nu}_{ab} \right\} C^{abc} 
\Phi_{\mu\nu}{}^c]
\end{eqnarray}
\begin{eqnarray}
\delta_0 \Phi_{\mu\nu}{}^a &=& D_{[\mu} z_{\nu]}{}^a + 
\eta_{\mu\nu}{}^a + \frac{4m_2}{(d-3)} e_{[\mu}{}^a \zeta_{\nu]},
\qquad \delta_0 \Omega_\mu{}^{abc} = D_\mu \eta^{abc} \nonumber \\
\delta_0 C_{\mu\nu} &=& D_{[\mu} \zeta_{\nu]} - 2 m_2 z_{[\mu\nu]},
\qquad \delta_1 C^{abc} = 6 m_2 \eta^{abc}
\end{eqnarray}
corresponds to massless representation of $dS$ group, while the other
one describes a so called partially massless spin 2 particle 
\cite{DW01,DW01a,DW01c,Zin01,SV06,DW06,Zin08b}.

\subsection{Gauge invariant objects}

Having in our disposal explicit form of the gauge transformations we
can construct gauge invariant curvatures for all eight fields (both
physical and auxiliary):
\begin{eqnarray}
{\cal R}_{\mu\nu}{}^{abc} &=& D_{[\mu} \Omega_{\nu]}{}^{abc} +
\frac{4m_1}{3(d-3)} e_{[\mu}{}^{[a} \omega_{\nu]}{}^{bc]} + 
\frac{4m_2}{3(d-3)} e_{[\mu}{}^{[a} C_{\nu]}{}^{bc]} \nonumber \\
{\cal T}_{\mu\nu\alpha}{}^a &=& D_{[\mu} \Phi_{\nu\alpha]}{}^a -
\Omega_{[\mu,\nu\alpha]}{}^a + \frac{2m_1}{3(d-3)} e_{[\mu}{}^a
f_{\nu,\alpha]}  + \frac{4m_2}{(d-3)} e_{[\mu}{}^a C_{\nu\alpha]}
\nonumber \\
{\cal F}_{\mu\nu}{}^{ab} &=& D_{[\mu} \omega_{\nu]}{}^{ab} + 2 m_1
\Omega_{[\mu,\nu]}{}^{ab} - \frac{2\tilde{m}_2}{(d-2)}  
e_{[\mu}{}^{[a} B_{\nu]}{}^{b]} \nonumber \\
T_{\mu\nu}{}^a &=& D_{[\mu} f_{\nu]}{}^a - \omega_{[\mu,\nu]}{}^a - 4
m_1 \Phi_{\mu\nu}{}^a + \frac{4\tilde{m}_2}{(d-2)} e_{[\mu}{}^a
B_{\nu]} \nonumber \\
{\cal C}_\mu{}^{abc} &=& D_\mu C^{abc} - 6 m_2 \Omega_\mu{}^{abc} - 
\frac{2\tilde{m}_1}{(d-2)} e_\mu{}^{[a} B^{bc]}  \\
{\cal C}_{\mu\nu\alpha} &=& D_{[\mu} C_{\nu\alpha]} - C_{\mu\nu\alpha}
+ 2 m_2 \Phi_{[\mu\nu,\alpha]} \nonumber \\
{\cal B}_\mu{}^{ab} &=& D_\mu B^{ab} + 4 \tilde{m}_2 \omega_\mu{}^{ab}
+ \frac{4\tilde{m}_1}{3} C_\mu{}^{ab} \nonumber \\
{\cal B}_{\mu\nu} &=& D_{[\mu} B_{\nu]} - B_{\mu\nu} - 2\tilde{m}_2
f_{[\mu,\nu]} - 4 \tilde{m}_1 C_{\mu\nu} \nonumber
\end{eqnarray}

Similarly to massive spin 2 case, if we consider the most general
Lagrangian quadratic in these curvatures and require that all higher
derivatives terms were absent and kinetic terms were diagonal, we will
still have some ambiguity in the choice of coefficients. In what
follows we will use the following concrete form of free Lagrangian
(see below):
\begin{eqnarray}
{\cal L}_0 &=& \epfour [ 
a_1 {\cal R}_{\mu\nu}{}^{abe} {\cal R}_{\alpha\beta}{}^{cde} +
a_2 {\cal F}_{\mu\nu}{}^{ab} {\cal F}_{\alpha\beta}{}^{cd} ] +
\eptwo [ a_3 {\cal C}_\mu{}^{acd} {\cal C}_\nu{}^{bcd} +
a_4 {\cal B}_\mu{}^{ac} {\cal B}_\nu{}^{bc} ] + \nonumber \\
 && + b_1 \epfour {\cal R}_{\mu\nu}{}^{abc} T_{\alpha\beta}{}^d 
+ b_2 \epthree {\cal C}_\mu{}^{abc} {\cal B}_{\nu\alpha}
\end{eqnarray}
$$
a_1 = - \frac{9}{512m_1{}^2}, \qquad
a_2 = \frac{2}{3(d-3)} a_1 = - \frac{3}{256(d-3)m_1{}^2}
$$
$$
a_3 = \frac{32(d-4)}{9(d-3)} a_1 = - \frac{(d-4)}{16(d-3)m_1{}^2},
\qquad a_4 = \frac{4(d-3)}{(d-2)} a_2 = - \frac{3}{64(d-2)m_1{}^2}
$$
$$
b_1 = - \frac{1}{32m_1}, \qquad b_2 = \frac{1}{24\tilde{m}_1}
$$

\subsection{Deformations}

Our next task is to construct appropriate deformations for all gauge
invariant objects. As in the case of massive spin 2 particle all we
have to do with curvatures for mixed symmetry tensor is to replace
$(A)dS$ covariant derivatives with the fully covariant ones. Using
explicit form of free gauge transformations and free curvatures it is
straightforward to find appropriate deformation for e/m field
strength. Indeed, let us consider the following ansatz for such
deformation:
$$
\hat{F}_{\mu\nu} = a_0 \varepsilon^{ij} [ \Omega_{[\mu}{}^{abc,i}
\Omega_{\nu]}{}^{abc,j} + \alpha_1 \omega_{[\mu}{}^{ab,i}
\omega_{\nu]}{}^{ab,j} + \alpha_2 C_{[\mu}{}^{ab,i} C_{\nu]}{}^{ab,j}
+ \alpha_3 B_{[\mu}{}^{a,i} B_{\nu]}{}^{a,j} ]
$$
Its variation under the $\eta^{abc}$ transformations has the form:
$$
\delta \hat{F}_{\mu\nu} = 2a_0 \varepsilon^{ij} [ - D_{[\mu}
\eta^{abc,j} \Omega_{\nu]}{}^{abc,i} - 2m_1\alpha_1 
\omega_{[\mu}{}^{ab,i} \eta_{\nu]}{}^{ab,j} + 6m_2\alpha_2 
C_{[\mu}{}^{ab,i} \eta_{\nu]}{}^{ab,j} ]
$$
Introducing appropriate correction:
$$
\delta A_\mu = 2a_0 \varepsilon^{ij} \Omega_\mu{}^{abc,i} \eta^{abc,j}
$$
we obtain:
$$
\delta \hat{F}_{\mu\nu} = 2a_0 \varepsilon^{ij} [ D_{[\mu}
 \Omega_{\nu]}{}^{abc,i} \eta^{abc,j} - 2m_1\alpha_1 
\omega_{[\mu}{}^{ab,i} \eta_{\nu]}{}^{ab,j} + 6m_2\alpha_2 
C_{[\mu}{}^{ab,i} \eta_{\nu]}{}^{ab,j} ]
$$
Comparing this expression with
$$
2a_0 \varepsilon^{ij} {\cal R}_{\mu\nu}{}^{abc,i} \eta^{abc,j} = 2a_0
\varepsilon^{ij} [ D_{[\mu} \Omega_{\nu]}{}^{abc,i} + 
\frac{4m_1}{(d-3)} e_{[\mu}{}^a \omega_{\nu]}{}^{bc,i} +
\frac{4m_2}{(d-3)} e_{[\mu}{}^a C_{\nu]}{}^{bc,i} ] \eta^{abc,j}
$$
we see that we have to put
$$
\alpha_1 = \frac{2}{(d-3)}, \qquad \alpha_2 = - \frac{2}{3(d-3)}
$$
Similarly, variation under the $\eta^{ab}$ transformations:
$$
\delta \hat{F}_{\mu\nu} = 2a_0 \varepsilon^{ij} [ - \alpha_1 D_{[\mu}
\eta^{ab,j} \omega_{\nu]}{}^{ab,i} + \frac{4m_1}{(d-3)} 
\Omega_{[\mu}{}^{abc,i} e_{\nu]}{}^a \eta^{bc,j} - 4\tilde{m}_2
\alpha_3 B_{[\mu}{}^{a,i} \eta_{\nu]}{}^{a,j} ]
$$
together with the following correction
$$
\delta A_\mu = 2\alpha_1a_0 \varepsilon^{ij} \omega_\mu{}^{ab,i}
\eta^{ab,j}
$$
give us
$$
\delta \hat{F}_{\mu\nu} = 2a_0 \varepsilon^{ij} [ \alpha_1 D_{[\mu}
 \omega_{\nu]}{}^{ab,i} \eta^{ab,j} + \frac{4m_1}{(d-3)} 
\Omega_{[\mu}{}^{abc,i} e_{\nu]}{}^a \eta^{bc,j} - 4\tilde{m}_2
\alpha_3 B_{[\mu}{}^{a,i} \eta_{\nu]}{}^{a,j} ]
$$
Comparing this expression with
$$
2\alpha_1a_0 \varepsilon^{ij} {\cal F}_{\mu\nu}{}^{ab,i} \eta^{ab,j} =
2\alpha_1a_0 \varepsilon^{ij} [ D_{[\mu} \omega_{\nu]}{}^{ab,i} + 2m_1
\Omega_{[\mu,\nu]}{}^{ab,i} - \frac{4\tilde{m}_2}{(d-2)} e_{[\mu}{}^a
B_{\nu]}{}^{b,i} ] \eta^{ab,j}
$$
we obtain
$$
\alpha_3 = - \frac{2}{(d-2)(d-3)}
$$
Thus we obtain finally
\begin{eqnarray}
\hat{F}_{\mu\nu} &=& F_{\mu\nu} + a_0 \varepsilon^{ij} [
\Omega_{[\mu}{}^{abc,i} \Omega_{\nu]}{}^{abc,j} 
+ \frac{2}{(d-3)} \omega_{[\mu}{}^{ab,i} \omega_{\nu]}{}^{ab,j} -
\nonumber \\
 && \qquad \qquad - \frac{2}{3(d-3)} C_{[\mu}{}^{ab,i} 
C_{\nu]}{}^{ab,j} - \frac{1}{2(d-3)(d-2)} B_{[\mu}{}^{a,i} 
B_{\nu]}{}^{a,j} ]
\end{eqnarray}
In this, if we supplement this deformation with the following
corrections to e/m field transformations:
\begin{equation}
\delta A_\mu = 2a_0 \varepsilon^{ij} [ \Omega_\mu{}^{abc,i}
\eta^{abc,j} + \frac{2}{(d-3)} \omega_\mu{}^{ab,i} \eta^{ab,j} ]
\end{equation}
then under massive mixed tensor gauge transformations we obtain:
\begin{equation}
\delta \hat{F}_{\mu\nu} = 2a_0 \varepsilon^{ij} [ 
{\cal R}_{\mu\nu}{}^{abc,i} \eta^{abc,j} + \frac{2}{(d-3)}
{\cal F}_{\mu\nu}{}^{ab,i} \eta^{ab,j} ]
\end{equation}

Thus, following general procedure, we consider the following
Lagrangian:
\begin{eqnarray}
{\cal L} &=& \epfour [ 
a_1 \hat{\cal R}_{\mu\nu}{}^{abe} \hat{\cal R}_{\alpha\beta}{}^{cde} +
a_2 \hat{\cal F}_{\mu\nu}{}^{ab} \hat{\cal F}_{\alpha\beta}{}^{cd} ] +
\eptwo [ a_3 \hat{\cal C}_\mu{}^{acd} \hat{\cal C}_\nu{}^{bcd} +
a_4 \hat{\cal B}_\mu{}^{ac} \hat{\cal B}_\nu{}^{bc} ] + \nonumber \\
 && + b_1 \epfour \hat{\cal R}_{\mu\nu}{}^{abc}
\hat{T}_{\alpha\beta}{}^d + \epthree b_2 \hat{\cal C}_\mu{}^{abc} 
\hat{\cal B}_{\nu\alpha} - \frac{1}{4} \hat{F}_{\mu\nu}{}^2
\end{eqnarray}

\subsection{Gauge invariance}

Now we have to calculate variations of this Lagrangian in the first
non-trivial approximation and to adjust parameters so that all such
variations vanish on-shell. This in turn would imply that they can be
compensated by appropriate corrections to gauge transformations.

$\eta^{abc}$ {\bf transformations}. In this case we obtain the
following variations of our Lagrangian:
\begin{eqnarray}
\delta {\cal L} &=&
- a_0 \varepsilon^{ij} F^{\mu\nu} {\cal R}_{\mu\nu}{}^{cde,i}
\eta^{cde,j} + 8 a_1 e_0 \varepsilon^{ij} {\cal R}_{\mu\nu}{}^{cde,i}
F^{cd} \eta^{\mu\nu e,j} + \nonumber \\
 && + 4 a_1 e_0 \varepsilon^{ij} \eptwo [ - 4 
{\cal R}_{\mu\nu}{}^{ace,i} \eta^{bde,j} + {\cal R}_{\mu\nu}{}^{abe,i}
\eta^{cde,j} ] F^{cd} + \nonumber \\
 && + 6 b_1 e_0 \varepsilon^{ij} \eptwo [ T_{\mu\nu}{}^{a,i}
\eta^{bcd,j} + T_{\mu\nu}{}^{c,i} \eta^{abd,j} ] F^{cd}
\end{eqnarray}
All terms in the second and the third lines clearly vanish on-shell
being proportional to $\Phi_{\mu\nu}{}^a$ and $\omega_\mu{}^{ab}$
equations correspondingly. Now using explicit expressions for free
curvatures it is not hard to check that the following identity holds:
\begin{equation}
{\cal R}_{[\mu\nu,\alpha\beta]}{}^a = 
- D_{[\mu}  {\cal T}_{\nu\alpha\beta]}{}^a 
- \frac{2m_1}{3(d-3)} e_{[\mu}{}^a T_{\nu\alpha,\beta]}
 - \frac{4m_2}{(d-3)} e_{[\mu}{}^a {\cal C}_{\nu\alpha\beta]}
\end{equation}
Thus ${\cal R}_{[\mu\nu,\alpha\beta]}{}^a$ vanish on-shell, in this
the following consequence appears:
$$
{\cal R}_{[\mu\nu,\alpha\beta]}{}^a = 0 \qquad \Longrightarrow \qquad
\eta^{abc} F^{de} {\cal R}_{ab,cde} = \frac{1}{3} F^{ab} \eta^{cde}
{\cal R}_{ab,cde}
$$
Thus if we put $a_0 = \frac{8e_0a_1}{3}$ then all variations vanish
on-shell and can be compensated with the corresponding corrections to
gauge transformations. Again we restrict ourselves with the
corrections for physical fields only which have the following form:
\begin{equation}
\delta_1 \Phi_{\mu\nu}{}^{a,i} = 2 a_0 \varepsilon^{ij}  [ 2 
F_{[\mu}{}^b \eta_{\nu]}{}^{ab,j} - \frac{1}{(d-3)} e_{[\mu}{}^a 
(F \eta)_{\nu]}{}^j ]
\end{equation}

$\eta^{ab}$ {\bf transformations}. Here we obtain the following
variations:
\begin{eqnarray*}
\delta_\eta {\cal L} &=& - 2 a_2 e_0 \varepsilon^{ij}
\eptwo [ 4 {\cal F}_{\mu\nu}{}^{ac,i} \eta^{bd,j} - 
{\cal F}_{\mu\nu}{}^{ab,i} \eta^{cd,j} ] F^{cd} - \\
 && - \frac{2a_0}{(d-3)} \varepsilon^{ij} F^{\mu\nu} 
{\cal F}_{\mu\nu}{}^{ab,i} \eta^{ab,j} + 8 a_2 e_0 \varepsilon^{ij} 
{\cal F}_{\mu\nu}{}^{ab,i} F^{ab} \eta^{\mu\nu,j}
\end{eqnarray*}
Terms in the first line vanish on-shell being proportional to free
$f_\mu{}^a$ equations. As for the second line, it is not hard to check
that the following identity holds:
\begin{equation}
{\cal F}_{[\mu\nu,\alpha]}{}^a = - D_{[\mu} T_{\nu\alpha]}{}^a -
4m_1 {\cal T}_{\mu\nu\alpha}{}^a - \frac{4\tilde{m}_2}{(d-2)} 
e_{[\mu}{}^a {\cal B}_{\nu\alpha]}
\end{equation}
This means that ${\cal F}_{[\mu\nu,\alpha]}{}^a$ vanish on-shell.
Thus if we put $a_0 = 4 a_2 e_0 (d-3)$ then all variations vanish 
on-shell. Comparing this relation with $a_0 = \frac{8e_0a_1}{3}$ we
see that $a_2 = \frac{2a_1}{3(d-3)}$ and this explains out choice of
concrete form for the free Lagrangian. Corrections to gauge
transformations for physical fields look as follows:
\begin{equation}
\delta f_\mu{}^{a,i} = \frac{2a_0}{(d-3)} \varepsilon^{ij} [ 2 
F_\mu{}^b \eta^{ab,j} - \frac{1}{(d-2)} e_\mu{}^a (F \eta)^j ]
\end{equation}

$\xi^a$ and $\Lambda$ {\bf transformations}. In these cases we obtain
simply:
\begin{equation}
\delta {\cal L} = b_1 e_0 \varepsilon^{ij} \epfour 
{\cal R}_{\mu\nu}{}^{abc,i} F_{\alpha\beta} \xi^{d,j} + b_2 e_0
\varepsilon^{ij} \epthree {\cal C}_\mu{}^{abc,i} F_{\nu\alpha}
\Lambda^j
\end{equation}
and these variations can be compensated by:
\begin{equation}
\delta_2 \Phi_{\mu\nu}{}^{a,i} = - 8b_1 e_0 \varepsilon^{ij}
F_{\mu\nu} \xi^{a,j}, \qquad 
\delta C_{\mu\nu}{}^i = 6b_2 e_0 \varepsilon^{ij} F_{\mu\nu}
\Lambda^j
\end{equation}

\subsection{Results}

Thus we have seen that the following Lagrangian:
\begin{eqnarray}
{\cal L} &=& \epfour [ 
a_1 \hat{\cal R}_{\mu\nu}{}^{abe} \hat{\cal R}_{\alpha\beta}{}^{cde} +
a_2 \hat{\cal F}_{\mu\nu}{}^{ab} \hat{\cal F}_{\alpha\beta}{}^{cd} ] +
\eptwo [ a_3 \hat{\cal C}_\mu{}^{acd} \hat{\cal C}_\nu{}^{bcd} +
a_4 \hat{\cal B}_\mu{}^{ac} \hat{\cal B}_\nu{}^{bc} ] + \nonumber \\
 && + b_1 \epfour \hat{\cal R}_{\mu\nu}{}^{abc}
\hat{T}_{\alpha\beta}{}^d + \epthree b_2 \hat{\cal C}_\mu{}^{abc} 
\hat{\cal B}_{\nu\alpha} - \frac{1}{4} \hat{F}_{\mu\nu}{}^2
\end{eqnarray}
is indeed gauge invariant in the linear approximation provided we
supplement it with the following corrections to gauge transformations:
\begin{eqnarray}
\delta A_\mu &=& 2a_0 \varepsilon^{ij} [ 
\Omega_\mu{}^{abc,i} \eta^{abc,j} + \frac{2}{(d-3)} 
\omega_\mu{}^{ab,i} \eta^{ab,j} ] \nonumber \\
\delta \Phi_{\mu\nu}{}^{a,i} &=& 2 a_0 \varepsilon^{ij}  [ 2 
F_{[\mu}{}^b \eta_{\nu]}{}^{ab,j} - \frac{1}{(d-3)} e_{[\mu}{}^a 
(F \eta)_{\nu]}{}^j ] - 8b_1 e_0 \varepsilon^{ij}
F_{\mu\nu} \xi^{a,j} \nonumber \\
\delta f_\mu{}^{a,i} &=& \frac{2a_0}{(d-3)} \varepsilon^{ij} [ 2 
F_\mu{}^b \eta^{ab,j} - \frac{1}{(d-2)} e_\mu{}^a (F \eta)^j ] \\
\delta C_{\mu\nu}{}^i &=& 6b_2 e_0 \varepsilon^{ij} F_{\mu\nu}
\Lambda^j \nonumber
\end{eqnarray}

As we have seen from the formulas above, electric charge $e_0 \sim a_0
m_1$ so it is impossible to take (partially) massless limit $m_1 \to
0$ in de Sitter space without switching off minimal e/m interactions.
In this, nothing prevent us from considering (partially) massless
limit $m_2 \to 0$ in anti de Sitter space. In this limit we obtain the
following simple Lagrangian:
\begin{equation}
{\cal L} = \epfour [ 
a_1 \hat{\cal R}_{\mu\nu}{}^{abe} \hat{\cal R}_{\alpha\beta}{}^{cde} +
a_2 \hat{\cal F}_{\mu\nu}{}^{ab} \hat{\cal F}_{\alpha\beta}{}^{cd} +
b_1 \hat{\cal R}_{\mu\nu}{}^{abc} \hat{T}_{\alpha\beta}{}^d ] -
\frac{1}{4} \hat{F}_{\mu\nu}{}^2
\end{equation}
where:
\begin{eqnarray}
\hat{F}_{\mu\nu} &=& F_{\mu\nu} + a_0 \varepsilon^{ij} [
\Omega_{[\mu}{}^{abc,i} \Omega_{\nu]}{}^{abc,j} 
+ \frac{2}{(d-3)} \omega_{[\mu}{}^{ab,i} \omega_{\nu]}{}^{ab,j} ]
\nonumber \\
{\cal R}_{\mu\nu}{}^{abc} &=& D_{[\mu} \Omega_{\nu]}{}^{abc} +
\frac{4m_1}{3(d-3)} e_{[\mu}{}^{[a} \omega_{\nu]}{}^{bc]} \nonumber \\
{\cal F}_{\mu\nu}{}^{ab} &=& D_{[\mu} \omega_{\nu]}{}^{ab} + 2 m_1
\Omega_{[\mu,\nu]}{}^{ab}  \\
T_{\mu\nu}{}^a &=& D_{[\mu} f_{\nu]}{}^a - \omega_{[\mu,\nu]}{}^a - 4
m_1 \Phi_{\mu\nu}{}^a \nonumber
\end{eqnarray}

\section*{Conclusion}

Thus we have seen that using frame-like gauge invariant formulation it
is indeed possible to extend Fradkin-Vasiliev procedure to the case of
electromagnetic interactions of massive particles. We constructed two
explicit examples: spin 2 and simplest mixed symmetry tensor. These
two cases are just the simplest representatives of two general class
of fields, namely completely symmetric and mixed symmetry ones, and it
is clear that the results obtained admit straightforward
generalization to higher spins as well.

\vskip 1cm \noindent
{\bf Acknowledgment}  \\
Author is grateful to N. Boulanger, E. Skvortsov and M. Vasiliev for
many useful discussions.


\newpage

\begin{thebibliography}{10}

\bibitem{BBS10}
X.~Bekaert, N.~Boulanger, P.~Sundell
{\it "How higher-spin gravity surpasses the spin two barrier: no-go
theorems versus yes-go examples",} arXiv:1007.0435.

\bibitem{FV87}
E.~S. Fradkin, M.~A. Vasiliev
{\it "On the gravitational interaction of massless higher-spin
fields",}
Phys. Lett. {\bf B189} (1987) 89.

\bibitem{FV87a}
E.~S. Fradkin, M.~A. Vasiliev
{\it "Cubic interaction in extended theories of massless higher-spin
fields",}
Nucl. Phys. {\bf B291} (1987) 141.

\bibitem{Zin08}
Yu.~M. Zinoviev
{\it "On spin 3 interacting with gravity",}
Class. Quantum Grav. {\bf 26} (2009) 035022, arXiv:0805.2226.

\bibitem{BLS08}
N.~Boulanger, S.~Leclercq, P.~Sundell
{\it "On The Uniqueness of Minimal Coupling in Higher-Spin Gauge
Theory",}
JHEP {\bf 0808} (2008) 056, arXiv:0805.2764.

\bibitem{Zin08a}
Yu.~M. Zinoviev
{\it "On spin 2 electromagnetic interactions",}
Mod. Phys. Lett. {\bf A24} (2009) 17, arXiv:0806.4030.

\bibitem{Tar10}
M.~Taronna
{\it "Higher Spins and String Interactions",} arXiv:1005.3061.

\bibitem{ST10}
A.~Sagnotti, M.~Taronna
{\it "String Lessons for Higher-Spin Interactions",}
Nucl. Phys. {\bf B842} (2011) 299, arXiv:1006.5242.

\bibitem{Sch10}
O.~Schlotterer
{\it "Higher Spin Scattering in Superstring Theory",} arXiv:1011.1235.

\bibitem{BCNS02}
C.~Bizdadea, E.~M. Cioroianu, I.~Negru, S.~O. Saliu
{\it "Lagrangian interactions within a special class of covariant
  mixed-symmetry type tensor gauge fields",}
Eur. Phys. J. {\bf C27} (2003) 457, arXiv:hep-th/0211158.

\bibitem{BBH02}
X.~Bekaert, N.~Boulanger, M.~Henneaux
{\it "Consistent deformations of dual formulations of linearized
gravity: A no-go result",}
Phys. Rev. {\bf D67} (2003) 044010, arXiv:hep-th/0210278.

\bibitem{BCCSS03}
C.~Bizdadea, C.~C. Ciobirca, E.~M. Cioroianu, S.~O. Saliu, S.~C.
Sararu
{\it "Interacting mixed-symmetry type tensor gauge fields of degrees
two and three: a four-dimensional cohomological approach",} 
arXiv:hep-th/0303079.

\bibitem{BCCSS03a}
C.~Bizdadea, C.~C. Ciobirca, E.~M. Cioroianu, S.~O. Saliu, S.~C.
Sararu
{\it "Interactions of a massless tensor field with the mixed symmetry
of the Riemann tensor. No-go results",}
Eur.Phys.J. {\bf C36} (2004) 253-270, arXiv:hep-th/0306154.

\bibitem{Sal03}
S.~O. Saliu
{\it "Consistent interactions in a three-dimensional theory with
tensor gauge fields of degrees two and three",}
Int. J. Mod. Phys. {\bf A18} (2003) 4451, arXiv:hep-th/0304121.

\bibitem{BC04}
N.~Boulanger, S.~Cnockaert
{\it "Consistent deformations of [p,p]-type gauge field theories",}
JHEP {\bf 0403} (2004) 031, arXiv:hep-th/0402180.

\bibitem{BBC04}
X.~Bekaert, N.~Boulanger, S.~Cnockaert
{\it "No Self-Interaction for Two-Column Massless Fields",}
J. Math. Phys. {\bf 46} (2005) 012303, arXiv:hep-th/0407102.

\bibitem{CS07}
C.~C. Ciobirca, S.~O. Saliu
{\it "Generalized couplings between an Abelian $p$-form and a (3,1)
mixed symmetry tensor field",} arXiv:hep-th/0702018.

\bibitem{BCNS07}
C.~Bizdadea, C.~C. Ciobirca, I.~Negru, S.~O. Saliu
{\it "Couplings between a single massless tensor field with the mixed
symmetry (3,1) and one vector field",}
Phys. Rev. {\bf D74} (2006) 045031, arXiv:0705.1048.

\bibitem{BCCS07}
C.~Bizdadea, C.~C. Ciobirca, E.~M. Cioroianu, S.~O. Saliu
{\it "Interactions between a massless tensor field with the mixed
symmetry of the Riemann tensor and a massless vector field",}
J. Phys. A: Math. Gen. {\bf 39} (2006) 10549-10564, arXiv:0705.1054.

\bibitem{BCS09}
C.~Bizdadea, D.~Cornea, S.~O. Saliu
{\it "No cross-interactions among different tensor fields with the
mixed symmetry (3,1) intermediated by a vector field",}
J. Phys. A Math. Theor. {\bf 41} (2008) 285202, arXiv:0901.4059.

\bibitem{BCDISS09}
C.~Bizdadea, E.~M. Cioroianu, A.~Danehkar, M.~Iordache, S.~O. Saliu,
S.~C. Sararu
{\it "Consistent interactions of dual linearized gravity in D=5:
couplings with a topological BF model",}
Eur. Phys. J. {\bf C63} (2009) 491-519, arXiv:0908.2169.

\bibitem{FM95}
E.~S. Fradkin, R.~R. Metsaev
{\it "Cubic scattering amplitudes for all massless representations of
the Poincare group in any space-time dimension",}
Phys. Rev. {\bf D52} (1995) 4660.

\bibitem{Alk10}
K.~B. Alkalaev
{\it "FV-type action for AdS(5) mixed-symmetry fields",}
arXiv:1011.6109.

\bibitem{Vas96}
M.~A. Vasiliev
{\it "Higher-Spin Gauge Theories in Four, Three and Two Dimensions",}
Int. J. Mod. Phys. {\bf D5} (1996) 763, arXiv:hep-th/9611024.

\bibitem{Vas01}
M.~A. Vasiliev
{\it "Cubic Interactions of Bosonic Higher Spin Gauge Fields in
$AdS_5$",}
Nucl.Phys. {\bf B616} (2001) 106-162; Erratum-ibid. B652 (2003) 407,
  arXiv:hep-th/0106200.

\bibitem{AV02}
K.~B. Alkalaev, M.~A. Vasiliev
{\it "N=1 Supersymmetric Theory of Higher Spin Gauge Fields in AdS(5)
at the Cubic Level",}
Nucl.Phys. {\bf B655} (2003) 57-92, arXiv:hep-th/0206068.

\bibitem{Vas80}
M.~A. Vasiliev
{\it "'Gauge' form of description of massless fields with arbitrary
spin",}
Sov. J. Nucl. Phys. {\bf 32} (1980) 439.

\bibitem{LV88}
V.~E. Lopatin, M.~A. Vasiliev
{\it "Free massless bosonic fields of arbitrary spin in d-dimensional
de sitter space",}
Mod. Phys. Lett. {\bf A3} (1988) 257.

\bibitem{Vas88}
M.~A. Vasiliev
{\it "Free massless fermionic fields of arbitrary spin in
d-dimensional de sitter space",}
Nucl. Phys. {\bf B301} (1988) 26.

\bibitem{Zin08b}
Yu.~M. Zinoviev
{\it "Frame-like gauge invariant formulation for massive high spin
particles",}
Nucl. Phys. {\bf B808} (2009) 185, arXiv:0808.1778.

\bibitem{PV10}
D.~S. Ponomarev, M.~A. Vasiliev
{\it "Frame-Like Action and Unfolded Formulation for Massive
Higher-Spin Fields",}
Nucl. Phys. {\bf B839} (2010) 466, arXiv:1001.0062.

\bibitem{Zin03}
Yu.~M. Zinoviev
{\it "First Order Formalism for Mixed Symmetry Tensor Fields",}
  arXiv:hep-th/0304067.

\bibitem{ASV03}
K.~B. Alkalaev, O.~V. Shaynkman, M.~A. Vasiliev
{\it "On the Frame-Like Formulation of Mixed-Symmetry Massless Fields
in (A)dS(d)",}
Nucl. Phys. {\bf B692} (2004) 363, arXiv:hep-th/0311164.

\bibitem{Alk03}
K.~B. Alkalaev
{\it "Two-column higher spin massless fields in AdS(d)",}
Theor. Math. Phys. {\bf 140} (2004) 1253, arXiv:hep-th/0311212.

\bibitem{ASV05}
K.~B. Alkalaev, O.~V. Shaynkman, M.~A. Vasiliev
{\it "Lagrangian Formulation for Free Mixed-Symmetry Bosonic Gauge
Fields in (A)dS(d)",}
JHEP {\bf 0508} (2005) 069, arXiv:hep-th/0501108.

\bibitem{ASV06}
K.~B. Alkalaev, O.~V. Shaynkman, M.~A. Vasiliev
{\it "Frame-like formulation for free mixed-symmetry bosonic massless
  higher-spin fields in AdS(d)",} arXiv:hep-th/0601225.

\bibitem{Skv08a}
E.~D. Skvortsov
{\it "Mixed-Symmetry Massless Fields in Minkowski space Unfolded",}
JHEP {\bf 0807} (2008) 004, arXiv:0801.2268.

\bibitem{Skv08}
E.~D. Skvortsov
{\it "Frame-like Actions for Massless Mixed-Symmetry Fields in
Minkowski space",}
Nucl. Phys. {\bf B808} (2009) 569, arXiv:0807.0903.

\bibitem{SZ10}
E.~D. Skvortsov, Yu.~M. Zinoviev
{\it "Frame-like Actions for Massless Mixed-Symmetry Fields in
Minkowski space. Fermions",}
Nucl. Phys. {\bf B843} (2011) 559, arXiv:1007.4944.

\bibitem{Zin03a}
Yu.~M. Zinoviev
{\it "First Order Formalism for Massive Mixed Symmetry Tensor Fields
in Minkowski and (A)dS Spaces",} arXiv:hep-th/0306292.

\bibitem{Zin08c}
Yu.~M. Zinoviev
{\it "Towards frame-like gauge invariant formulation for massive mixed
symmetry bosonic fields",}
Nucl. Phys. {\bf B812} (2009) 46, arXiv:0809.3287.

\bibitem{Zin09b}
Yu.~M. Zinoviev
{\it "Frame-like gauge invariant formulation for mixed symmetry
fermionic fields",}
Nucl. Phys. {\bf B821} (2009) 21-47, arXiv:0904.0549.

\bibitem{Zin09c}
Yu.~M. Zinoviev
{\it "Towards frame-like gauge invariant formulation for massive mixed
symmetry bosonic fields. II. General Young tableau with two rows",}
Nucl. Phys. {\bf B826} (2010) 490, arXiv:0907.2140.

\bibitem{Zin09}
Yu.~M. Zinoviev
{\it "On massive spin 2 electromagnetic interactions",}
Nucl. Phys. {\bf B821} (2009) 431-451, arXiv:0901.3462.

\bibitem{DW01}
S.~Deser, A.~Waldron
{\it "Gauge Invariance and Phases of Massive Higher Spins in (A)dS",}
Phys. Rev. Lett. {\bf 87} (2001) 031601, arXiv:hep-th/0102166.

\bibitem{DW01a}
S.~Deser, A.~Waldron
{\it "Partial Masslessness of Higher Spins in (A)dS",}
Nucl. Phys. {\bf B607} (2001) 577, arXiv:hep-th/0103198.

\bibitem{DW01c}
S.~Deser, A.~Waldron
{\it "Null Propagation of Partially Massless Higher Spins in (A)dS and
  Cosmological Constant Speculations",}
Phys. Lett. {\bf B513} (2001) 137, arXiv:hep-th/0105181.

\bibitem{Zin01}
Yu.~M. Zinoviev
{\it "On Massive High Spin Particles in (A)dS",} arXiv:hep-th/0108192.

\bibitem{SV06}
E.~D. Skvortsov, M.~A. Vasiliev
{\it "Geometric Formulation for Partially Massless Fields",}
Nucl. Phys. {\bf B756} (2006) 117, arXiv:hep-th/0601095.

\bibitem{DW06}
S.~Deser, A.~Waldron
{\it "Partially Massless Spin 2 Electrodynamics",}
Phys. Rev. {\bf D74} (2006) 084036, arXiv:hep-th/0609113.

\bibitem{BMV00}
L.~Brink, R.~R. Metsaev, M.~A. Vasiliev
{\it "How massless are massless fields in $AdS_d$",}
Nucl. Phys. {\bf B586} (2000) 183, arXiv:hep-th/0005136.

\end{thebibliography}
\end{document}